\documentclass[12pt]{article}
\usepackage[top=1in, bottom=1in, left=1in, right=1in]{geometry}
\usepackage{graphicx}
\usepackage{amsmath,amssymb}
\usepackage{color}
\usepackage{hyperref}
\usepackage{slashed}
\usepackage{soul}
\usepackage[font=footnotesize]{caption}
\usepackage[dvipsnames]{xcolor} 

\newcommand{\matrixbb}[4]{\left(\hspace{-5 pt}\begin{tabular}{ c c } ${#1}$ & ${#2}$ \\ ${#3}$ & ${#4}$ \end{tabular}\hspace{-5 pt}\right)}
\newcommand{\matrixcc}[9]{\left(\hspace{-5 pt}\begin{tabular}{ c c c } ${#1}$ & ${#2}$& ${#3}$ \\ ${#4}$ & ${#5}$ & ${#6}$\\${#7}$& ${#8}$& ${#9}$\end{tabular}\hspace{-5 pt}\right)}
\newcommand{\matrixba}[2]{\left(\hspace{-5 pt}\begin{tabular}{ c } ${#1}$ \\ ${#2}$ \end{tabular}\hspace{-5 pt}\right)}
\newcommand{\matrixab}[2]{\left(\hspace{-5 pt}\begin{tabular}{ c c }${#1}$ & ${#2}$\end{tabular}\hspace{-5 pt}\right)}
\newcommand{\matrixca}[3]{\left(\hspace{-5 pt}\begin{tabular}{ c } ${#1}$ \\ ${#2}$ \\ ${#3}$ \end{tabular}\hspace{-5 pt}\right)}
\newcommand{\matrixac}[3]{\left(\hspace{-5 pt}\begin{tabular}{ c c c }${#1}$ & ${#2}$& ${#3}$\end{tabular}\hspace{-5 pt}\right)}

\definecolor{darkblue}{rgb}{0,0,0.9}
\hypersetup{
	linktocpage,
	colorlinks,
	citecolor=darkblue,
	filecolor=darkblue,
	linkcolor=darkblue,
	urlcolor=darkblue,
	pdftitle={Direct Search Implications for a Custodially-Embedded Composite Top},%
	pdfauthor={R. S. Chivukula, R. Foadi, D. Foren, and E. H. Simmons}
}

\title{Direct Search Implications for a Custodially-Embedded Composite Top}
\author{R. Sekhar Chivukula, Roshan Foadi, \\ Dennis Foren, and Elizabeth H. Simmons}
\date{}

\begin{document}
\maketitle
\vspace{-27 pt}
\begin{center}
{\it Department of Physics and Astronomy, Michigan State University,\\
East Lansing, MI, 48825, USA}
\end{center}

\abstract{We assess current experimental constraints on the bi-doublet + singlet model of top compositeness previously proposed in the literature. This model extends the standard model's spectrum by adding a custodially-embedded vector-like electroweak bi-doublet of quarks and a vector-like electroweak singlet quark. While either of those states alone would produce a model in tension with constraints from precision electroweak data, in combination they can produce a viable model.  We show that current precision electroweak data, in the wake of the Higgs discovery, accommodate the model and we explore the impact of direct collider searches for the partners of the top quark.}

\section{Introduction}

In the context of the Standard Model (SM), the top quark's large mass is an enduring mystery. All SM fermions are chiral: their left-handed components are weak doublets while their right-handed components are weak singlets. This property forbids SM fermion mass terms of the Dirac form, because they would violate the electroweak gauge symmetry. The Standard Model solves this dilemma via Yukawa interactions, which couple left-handed and right-handed components of fermions to the Higgs doublet. When the electroweak symmetry spontaneously breaks to electromagnetism and the Higgs doublet spontaneously acquires a nonzero vacuum expectation value, all of those Yukawa interactions yield mass terms for the fermions.  Hence, the SM implies a tree-level proportionality between the Yukawa coupling and mass of each fermion. The top quark's large mass ($173\text{ GeV}$) demands its Yukawa coupling be nearly one: $y_t \sim \sqrt{2}m_t/v \lesssim 1$, where $v=246\text{ GeV}$ is the electroweak scale. Consequently, the Yukawa coupling of the top quark is far larger than the Yukawa couplings of the other SM fermions.

The available evidence implies the SM is an excellent description of nature, including its hypothesis that all fermions are chiral.  However, there could exist heavier, undiscovered fermions that disrupt this pattern. In fact, electroweak precision data suggest there is more room in theories beyond the Standard Model for fermions that are vectorial under the electroweak gauge group than for new chiral fermions \cite{Agashe:2014kda}.  Notably, vector-like fermions can possess Dirac mass terms without violating the electroweak gauge symmetry; these Dirac masses are not tied to the size of the Higgs boson's vacuum expectation value or Yukawa couplings.

The existence of vector-like quarks may provide an alternative explanation for the top quark's large mass. Suppose there exists some vector-like gauge-eigenstate quark (with a Dirac mass) whose quantum numbers allow it to mix with a chiral gauge-eigenstate quark (with a Yukawa coupling to the Higgs boson). The mass eigenstates of the system would each be a superposition of the two gauge eigenstates and the masses would arise from a combination of Dirac and Yukawa terms.  If the top quark is composite in this manner, then only a small fraction of its mass need be due to Yukawa interactions while the majority could come from a Dirac mass term. The Yukawa coupling of the top quark then must be larger than the SM prediction to reproduce the measured mass of the top quark. As a result, we can attribute the size of $y_t$ to top compositeness and consider $y_t$ naturally large \cite{Dobrescu:1997nm,Chivukula:1998wd,Pomarol:2008bh}.

This work explores the possibility that vector-like fermions of this kind exist and play a role in the composition and phenomenology of the top quark\footnote{While the other quarks and leptons could also be superpositions of vector-like and chiral fermions, they are so light relative to the top quark that the degree of mixing would be far smaller. Furthermore, data significantly constrains the degree to which the couplings of most other fermions deviate from the SM values, severely limiting the degree of mixing \cite{Aad:2011yn}. Therefore we ignore possible compositeness of the light fermions.}.  Restricting the Higgs sector to consist of only the observed Higgs doublet constrains these vector-like ``top partner" states to be EW singlets, doublets, or triplets \cite{Aguila:2000}. In minimal effective theories, the triplet case generates phenomenologically interesting effects that are already included in the singlet and doublet cases \cite{Chivukula:2011}; hence we will focus on a model employing the singlet and doublet states directly.

If the physical top quark contains a substantial superposition of vectorial and chiral fermions, its couplings to other particles are altered from their SM predictions. This has immediate consequences for the electroweak precision data, which are heavily influenced by the top quark's couplings to electroweak gauge bosons. Models with composite top quarks are expected to deviate from the SM through the introduction of tree-level flavor-changing neutral currents, right-handed charged currents, and significant weak isospin violation. In particular, in the flavor-conserving sector, these models are significantly constrained by the oblique parameters $\hat{S}$ and $\hat{T}$ and corrections to the $Zb_L\overline{b}_L$ coupling $g_{Lb}$.

An analysis \cite{Chivukula:2011} of composite top models performed before the 2012 Higgs discovery \cite{ATLAS:2012}-\cite{CMS:2013} suggested that including a particular combination of new vector-like fermions could lead to a ``bi-doublet + singlet model" consistent with electroweak data.  In this paper, we revisit the model in light of the Higgs boson discovery and other data from LHC Run I.  We establish that the model remains viable, demonstrate how the new data has further constrained the open parameter space, and suggest the likely impact of further searches for new physics at the LHC.  

In section 2, we review the essential features of the bi-doublet + singlet model from Ref \cite{Chivukula:2011}. Section 3 details how this model influences the relevant experimental parameters. Section 4 presents a new analysis of the model that includes updated electroweak precision constraints, the measured value of the Higgs boson's mass, and the impact of direct searches for vector-like top partner fermions. Finally, section 5 summarizes our conclusions and addresses the implications of prospective LHC Run-II direct limits.

\begin{table}[t]
	\centering
	\renewcommand{\arraystretch}{1.5}
	\begin{tabular}{l | c | c | c}
	\hspace{5 pt}  & $T_{3}^f$ &$Y^f$ & $Q_f$\\
	\hline
	{\color{black}$t_{0L}$}  & $+\frac{1}{2}$ & $+\frac{1}{3}$&  $+\frac{2}{3}$\\
	{\color{black}$b_{0L}$} & $-\frac{1}{2}$ & $+\frac{1}{3}$& $-\frac{1}{3}$ \\
	\hline
	{\color{black} $t_{0R}$  }& $0$ & $+\frac{4}{3}$&  $+\frac{2}{3}$ \\
	\hline
	$b_{0R}$ & $0$ & $-\frac{2}{3}$& $-\frac{1}{3}$\\
	\hline
	{\color{black} $t_{1L},t_{1R}$ }  & $0$ & $+\frac{4}{3}$& $+\frac{2}{3}$\\
	\hline
	$t_{1L}^q,t_{1R}^q$ & $+\frac{1}{2}$ & $+\frac{1}{3}$ & $+\frac{2}{3}$\\
	$b_{1L},b_{1R}$ & $-\frac{1}{2}$ & $+\frac{1}{3}$ & $-\frac{1}{3}$\\
	\hline
	{\color{black} $\Omega_{1L},\Omega_{1R}$}  & $+\frac{1}{2}$ & $+\frac{7}{3}$& $+\frac{5}{3}$ \\
	{\color{black} $t^{\Psi}_{1L},t^{\Psi}_{1R}$}  & $-\frac{1}{2}$ & $+\frac{7}{3}$ & $+\frac{2}{3}$
	\end{tabular}
	\captionsetup{width=0.8\textwidth}
	\caption{The third-generation fermion field content of the bi-doublet + singlet model. The $SU(2)_L$, $U(1)_Y$, and $U(1)_Q$ quantum number of each field is listed, respectively, in the first, second, and third column.}
\end{table}

\section{The Model}
We will now review the essential features of the bi-doublet + singlet model \cite{Chivukula:2011}. The gauge sector is identical to that of the Standard Model, as are the Lagrangian terms for the leptons, the first and second generation quarks, and the Higgs potential. New physics enters through the top-quark sector, via the following Lagrangian terms:
\begin{align}
\mathcal{L}_{bi\text{-}doublet+singlet} &= \overline{q}_{0L}i\slashed{D}q_{0L}+\overline{t}_{0R}i\slashed{D}t_{0R}+\overline{b}_{0R}i\slashed{D}b_{0R}+\overline{t}_1i\slashed{D}t_1 +\text{Tr}\left(\overline{Q}_1i\slashed{D}Q_1\right) \nonumber \\
&\hspace{20 pt} - M_t\overline{t}_1t_1-M_q\text{Tr}\left(\overline{Q}_1Q_1\right) -\mu_q\left(\overline{q}_{0L}q_{1R}+\text{h.c.}\right)-\mu_t\left(\overline{t}_{1L}t_{0R}+\text{h.c.}\right) \nonumber\\
&\hspace{20 pt}-y_q\left[\text{Tr}\left(\overline{Q}_{1L}\Phi\right)t_{0R}+\text{h.c.}\right]-y_t\left[\text{Tr}\left(\overline{Q}_{1L}\Phi\right)t_{1R}+\text{h.c.}\right]
\end{align}
where
\begin{align}
Q_1\equiv \left(q_{1}, \Psi_{1}\right) \equiv \matrixbb{t_1^q}{\Omega_1}{b_1}{t^\Psi_1}\hspace{40 pt}\Phi \equiv \left(\tilde{\varphi},\varphi\right)\equiv\left(i\sigma_2\varphi^*,\varphi\right) \label{QPhi}
\end{align}
and $\varphi$ is the usual Higgs doublet. The field content of the top-sector Lagrangian is summarized in Table 1. The Lagrangian features a left-handed SM-like EW doublet $q_{0L}$ consisting of the SM top and bottom quarks $(t_{0L}, b_{0L})$, and right-handed SM-like EW singlets $t_{0R}$ and $b_{0R}$.  By comparison with the SM, this model possesses 5 new fields: three top-like fields ($t^q_1$, $t^\Psi_1$, $t_1$), a bottom-like field ($b_1$), and an exotic quark field with $+5/3$ charge ($\Omega_1$).

The field combinations $Q_1$ and $\Phi$ are constructed to (approximately) preserve a global $O(4)\sim SU(2)_L\times SU(2)_R\times P_{LR}$ symmetry. The $P_{LR}$ symmetry acts on the fields as follows:
\begin{align}
Q_1 \rightarrow -\epsilon Q_{1}^T \epsilon\hspace{23 pt}\Phi\rightarrow -\epsilon \Phi^T \epsilon\hspace{23 pt}q_{0L}\rightarrow q_{0L} \hspace{23 pt}t_{0R}\rightarrow t_{0R} \hspace{23 pt} t_1\rightarrow t_1
\end{align}
where $\epsilon$ is an $SU(2)$ parity operation,
\begin{align}
\epsilon \equiv \matrixbb{0}{1}{-1}{0} \label{Epsilon}
\end{align}
Qualitatively, we expect cancellations in corrections to the $Z\to b\bar{b}$ amplitude \cite{Agashe:2006at} between the $t^q_1$ and $t^\Psi_1$ fields because $t^q_1$ carries isospin $T_{L3}=\frac{1}{2}$ while $t^\Psi_1$ carries isospin $T_{L3}=-\frac{1}{2}$. Under the $P_{LR}$ transformation, most of the terms in the Lagrangian are invariant. Exceptions are the $\mu_q$ and $\mu_t$ terms, which impose soft breaking of the global $O(4)$ down to a global $O(3)\sim SU(2)_V \times P_{LR}$.

We assume the SM bottom Yukawa coupling satisfies $y_b v \ll \mu_q$ and $y_b v\ll M_q$ so we can neglect $y_b$ in our model. The four top-like fermions will mix with one another to produce four top-like physical states. Similarly, the two bottom-like fermions will mix with one another to produce two bottom-like physical states. The bottom-like states mix in the same way as described for the SM + EW-doublet model of Ref \cite{Chivukula:2011}. The exotic $+5/3$ fermion is unique in its quantum numbers and hence does not mix with any other particles.

The SM limit of the bi-doublet + singlet model results from taking $\mu_q\gg M_q\rightarrow\infty$ and $M_t=\mu_t\rightarrow \infty$. In this limit, the left-handed EW doublet $q_{1L}$ becomes the SM left-handed top and bottom quarks, while the EW singlets $t_{1R}$ and $b_{0R}$ become the SM right-handed top and bottom quarks. All of the other fields listed in Table 1 mix into heavy states that decouple from the remainder of the theory.

To preserve the same qualitative behavior with fewer degrees of freedom, we will follow the analysis in Ref \cite{Chivukula:2011} and consider the $\mu_q \rightarrow \infty$ limit of the bi-doublet + singlet model. This effectively eliminates $q_{0L}$ and $q_{1R}$, and reduces the top Lagrangian to
\begin{align}
\mathcal{L}_{top} &= \overline{t}_{0R}i\slashed{D}t_{0R}+\overline{t}_1i\slashed{D}t_1 +\text{Tr}\left(\overline{Q}_{1L}i\slashed{D}Q_{1L}\right)+\overline{\Psi}_{1R}i\slashed{D}\Psi_{1R}- M_t\overline{t}_1t_1\nonumber\\
&\hspace{8 pt} -M_q\left(\overline{\Psi}_{1L}\Psi_{1R}+\text{h.c.}\right)-\mu_t\left(\overline{t}_{1L}t_{0R}+\text{h.c.}\right)-y_t\left[\text{Tr}\left(\overline{Q}_{1L}\Phi\right)t_{1R}+\text{h.c.}\right]\label{eqLag}
\end{align}
With $q_{0L}$ removed, $q_{1L}$ now plays the role of the left-handed SM-like top-bottom weak doublet. Note that this is precisely the doublet-extended standard model (DESM) of Ref. \cite{Chivukula:2009} with an added vector-like EW singlet. As such, we will refer to this limit as the DESM + singlet model.

The DESM + singlet Lagrangian contains four parameters, one of which is fixed by demanding that the lightest top-like mass eigenstate correspond to the physical top quark. Furthermore, we can rearrange the other parameters into a more convenient combination, and define,
\begin{align}
\tan\beta \equiv \dfrac{\mu_t}{M_t}
\end{align}
such that our three free parameters are $\sin\beta$, $M_q$, and $M_t$. (Because $\mu_t$ and $M_t$ are positive, $\beta$ must be between $0$ and $\pi/2$, so that $\sin\beta$ ranges between $0$ and $1$.)

Electroweak symmetry breaking proceeds in the usual way,
\begin{align}
\varphi = \dfrac{1}{\sqrt{2}} \matrixba{\phi^1+i\phi^2}{\phi^0 +i\phi^3} \hspace{10 pt}\rightarrow \hspace{10 pt}\dfrac{1}{\sqrt{2}}\matrixba{0}{v}+\dfrac{1}{\sqrt{2}}\matrixba{\phi^1+i\phi^2}{H+i\phi^3}, \label{SingletLag}
\end{align}
where $H$ is the Higgs boson. The resulting mass terms for the third-generation quarks are,
\begin{align}
 \mathcal{L}_{top} &\supset -\matrixac{\overline{t}^q_{1L}}{\overline{t}^\Psi_{1L}}{\overline{t}_{1L}}\matrixcc{0}{0}{\hat{m}_t}{0}{M_q}{\hat{m}_t}{\mu_t}{0}{M_t}\matrixca{t_{0R}}{t^\Psi_{1R}}{t_{1R}}-M_q\overline{\Omega}_{1L}\Omega_{1R}\\
 &\hspace{20 pt}-{\lim_{\mu_q\rightarrow\infty}}\matrixab{\overline{b}_{0L}}{\overline{b}_{1L}}\matrixbb{0}{\mu_q}{0}{M_q}\matrixba{b_{0R}}{b_{1R}}+\text{h.c.},
\end{align}
where the quantity $\hat{m}_t \equiv y_t v/\sqrt{2}$ is the tree-level standard model prediction for the top quark mass (in the limit of negligible CKM mixing). Note that $\Omega=\Omega_1$ is a mass eigenstate with mass $M_q$. We diagonalize the top-related mass matrix in the usual way with unitary matrices $L_t$, $R_t$, defined by,
\begin{align*}
 \matrixca{{t}_{1L}^q}{{t}_{1L}^\Psi}{{t}_{1L}}\equiv L_t\matrixca{t_L}{T_L^\Psi}{T_L}\hspace{25 pt} \matrixca{t_{0R}}{t_{1R}^\Psi}{t_{1R}}\equiv R_t\matrixca{t_R}{T_R^\Psi}{T_R}
\end{align*}
where $t$ denotes the physical top quark and $T^\Psi$, $T$ are the physical top partners. We define $T$ to be the lighter of the two top partners. In the limit with  $M_t,M_q\gg m_t$, we find, to second order in $m_t$,
\begin{align}
 m_T = \min\left\{\dfrac{M_t}{\cos\beta},M_q\right\}\hspace{25 pt}m_\Psi = \max\left\{\dfrac{M_t}{\cos\beta},M_q\right\}
\end{align}where $m_T$ is the mass of the lighter top partner and $m_\Psi$ is the mass of the heavier top partner.

In the full bi-doublet + singlet model, diagonalization of the bottom-related mass matrix occurs according to,
\begin{align}
 \matrixba{b_{0L}}{b_{1L}}=\matrixbb{\cos\alpha}{\sin\alpha}{-\sin\alpha}{\cos\alpha}\matrixba{b_L}{B_L}\hspace{20 pt}\matrixba{b_{0R}}{b_{1R}}=\matrixba{b_R}{B_R} \label{BottomMixing}
\end{align}
where $\tan\alpha\equiv \mu_q/M_q$. In the DESM + singlet model, we take $\mu_q\rightarrow \infty$ (such that $\alpha\rightarrow \pi/2$). This drives $m_b\rightarrow 0$ and $m_B\rightarrow\infty$ \cite{Chivukula:2011}. Therefore, the DESM + singlet model predicts a massless bottom quark and an infinitely massive bottom partner that decouples from the theory.

The SM limit of the DESM + singlet model results from additionally taking $M_q\rightarrow\infty$, $M_t\rightarrow \infty$, and $\sin\beta\rightarrow 1/\sqrt{2}$. As we discuss in Section 3.1, the Standard Model predicts $g_{Lb}$ to be approximately two standard deviations below the experimentally measured value. Therefore, for the purposes of this analysis, the SM limit of the DESM + singlet model lies in disfavored regions of parameter space.

\section{Considering Precision Electroweak Constraints}
\subsection{Relevant experimental parameters}
Mixing among vector-like and chiral fermions leads to non-SM couplings of fermions to the electroweak gauge bosons. These altered couplings can induce sizable deviations from SM predictions. Of the SM quarks, the top quark is the least experimentally constrained, and so it presents us with the widest window for discovering new effects without violating current experimental limits \cite{Aad:2011yn}. This simultaneously affords an opportunity to explain the large size of the top quark mass.

We expect the extensions of the top sector discussed here to be constrained by the Peskin-Takeuchi parameters, $\hat{S}\equiv \hat{\alpha}(M_Z) S$ and $\hat{T}\equiv \hat{\alpha}(M_Z) T$ \cite{Peskin:1991sw}. Corrections to the $\hat{U}$-parameter due to new physics are anticipated to be small compared to those for $\hat{S}$ and $\hat{T}$. Furthermore, our model generates more significant corrections to $\hat{T}$ than to $\hat{S}$ because we only introduce vector-like fermions. The current experimental limits on $S$ and $T$ at the $Z$-pole corresponding to a Higgs boson with mass $m_H=125\text{ GeV}$ are $S=-0.03\pm 0.10$ and $T=0.01\pm 0.12$ \cite{Agashe:2014kda},\cite{ALEPH:2005ab}. Using $\hat{\alpha}(M_Z)^{-1}=127.940\pm0.014$, $M_Z=91.1876\pm 0.0021$ GeV, and $m_t=173.24\pm0.81$ GeV \cite{Agashe:2014kda}, we will compare our theoretical predictions against $\hat{S}$ and $\hat{T}$. Note that both measured values are consistent with the SM prediction $S=T=0$, so any content beyond the SM must generate largely self-canceling corrections.

The $Zb_L\overline{b}_L$ coupling $g_{Lb}$ can also receive large corrections when one augments the top sector. Experimentally, $g_{Lb}=-0.4182\pm0.0015$, while the SM predicts a value of $g_{Lb,SM}=-0.42114\pm\substack{0.00045 \\ 0.00024}$ \cite{ALEPH:2005ab}. Because the SM value sits approximately two standard deviations below the experimental value, experiment favors a positive value of  $\delta g_{Lb}\equiv g_{Lb}-g_{Lb,SM}\approx 0.003$. In order to compare our model with experiment, we have calculated $g_{Lb}$ in the gaugeless limit, as discussed in Ref \cite{Chivukula:2011}.

We are interested in models that  reproduce all electroweak precision data to within $1\sigma$ of the experimentally measured values. Consequently, the Standard Model is disfavored in our analysis.


\subsection{Applications to top partner models}

The simplest top partner models that add a single kind of vector-like quark to the SM spectrum struggle to simultaneously satisfy constraints from $\hat{T}$ and $g_{Lb}$. {Explicit analysis demonstrates that $\hat{S}$ provides a weaker constraint than $\hat{T}$; therefore we will discuss $\hat{S}$ no further. For instance, adding an EW vector-like singlet moves $\hat{T}$ and $g_{Lb}$ in the positive direction such that by the time $g_{Lb}$ is within $1\sigma$ of its experimental value, $\hat{T}$ is already well outside of the experimental $1\sigma$ band \cite{Chivukula:2011}. Incorporating an EW vector-like doublet, instead, yields a qualitatively similar conclusion, with even larger discrepancies. The addition of an EW vector-like bi-doublet (imbued with additional symmetry structures) generates negative corrections to $\hat{T}$ and preserves the SM value of $g_{Lb}$. Adding just one kind of vector-like quark to the SM spectrum does not produce a viable model.

However, Ref \cite{Chivukula:2011} constructed an experimentally consistent theory by combining vector-like bi-doublet and vector-like singlet extensions of the SM spectrum; this is the bi-doublet + singlet model described in Section 2. Relative to the separate EW bi-doublet extension and EW singlet extension, the bi-doublet + singlet model reproduces the measured $\hat{T}$ to within $1\sigma$ across a larger region of parameter space. This is accomplished through cancellations between positive corrections due to the EW singlet and negative corrections due to the EW bi-doublet. The bi-doublet + singlet model also controls positive $\delta g_{Lb}$ corrections by imposing symmetries on the bi-doublet.

Specifically, $\hat{T}$ corrections are diminished by imposing a global $SU(2)_L\times SU(2)_R$ symmetry on the symmetry-breaking sector and collapsing it to the usual custodial $SU(2)_c$ upon electroweak symmetry breaking \cite{Weinstein:1973gj},\cite{Sikivie:1980hm}. Agashe {\it et al.} noted that this symmetry could also protect $g_{Lb}$ by imposing an additional parity symmetry $P_{LR}$ between the $SU(2)_L$ and $SU(2)_R$ and demanding that $b_L$ be an eigenstate of $P_{LR}$ \cite{Agashe:2006at}.

The DESM + singlet limit of the bi-doublet + singlet model includes $b_L$ as an odd eigenstate of $P_{LR}$ with quantum numbers $T_L=T_R=1/2$ and $T_L^3=T_R^3=-1/2$ (e.g., $b_L$ is embedded in a bi-doublet $(2,2)_{2/3}$ of $SU(2)_L\times SU(2)_R\times U(1)_X$) so that its coupling to the $Z$-boson will receive suppressed corrections \cite{Agashe:2006at}. This is equivalent to embedding $b_L$ in the vector portion of $SO(3)\sim SU(2)_c$, such that $\delta g_{Lb} = \delta g_{Rb}$. Because $b_L$ is odd under $P_{LR}$, one finds that $\delta g_{Lb} = -\delta g_{Rb}$, and we conclude that $\delta g_{Lb}=\delta g_{Rb}=0$. This would only be approximately true in the full theory because the global symmetries are not exact. For example, the mass splitting $m_t\neq m_b$ breaks $SU(2)_c$ even in the limit of zero hypercharge. Regardless, embedding $b_L$ in an EW bi-doublet protects $g_{Lb}$ from large corrections. This allows the vector-like singlet quark to provide the primary corrections to $g_{Lb}$, making agreement with experiment feasible.

Figure 1 shows the shaded regions of parameter space that are excluded by present constraints on  $\hat{T}$ (coarse hatching) and $g_{Lb}$ (fine hatching), with the measured value of the Higgs mass taken into account; the white region in each pane of fixed $\sin\beta$ shows the area in the $M_t$ {\it v.s.} $M_q$ plane that remains viable in light of precision EW data.

\section{Incorporating Top Partner Searches}
Since 2011, the LHC experiments have performed a number of direct searches for top partners \cite{Agashe:2014kda}. Each search assumes that the top partners have particular decay properties and reports lower limits on their masses based on the absence of signal events. In particular, a top partner $T$ is often assumed to decay in only three ways: $T\rightarrow Zt$, $T\rightarrow Wb$, and $T\rightarrow Ht$. Since this assumption holds for the lighter top partner in our model, we can apply the existing search results to determine constraints on the top partner's mass.

There are three top-like physical quarks in the DESM + singlet model. By construction, the lightest of these mass eigenstates is the physical top quark. The remaining two mass eigenstates are top partners, where we defined $T$ ($T^\Psi$) to be the lighter (heavier) top partner. {The heavier top partner's mass varies between $0.80$ and $21$ TeV in regions of parameter space allowed by $\hat{T}$ and $g_{Lb}$, which overlaps direct limits; however, any point of parameter space for which the heavier top partner is eliminated by this constraint is already eliminated from direct search constraints on the lighter top partner.} Therefore, we focus on the constraints the data place upon the lighter top partner. The labeled curves within the white region of Figure 1 illustrate how $m_T$ varies as a function of $(M_t,M_q)$, and for different values of $\sin\beta$, across the regions allowed by the  electroweak precision data ($\hat{T}$, $\delta g_{Lb}$).

Different points in the $(M_q,M_t,\sin\beta)$ parameter space of our model generally correspond to different coupling values, and yield different values of $Br(T\rightarrow Zt)$, $Br(T\rightarrow Wb)$, and $Br(T\rightarrow Ht)$. Therefore, applying the experimental limits from direct searches requires careful analysis of $(M_q,M_t,\sin\beta)$ space.

{CMS and ATLAS performed inclusive vector-like top partner searches that incorporate all values of top partner branching fractions into $bW$, $tZ$, and $tH$ (assuming only these decay modes). The ATLAS search finds its strongest limit ($950$ GeV) when $Br(T\rightarrow Ht)\approx1$ \cite{ATLAS:2015dka}. Because the DESM + singlet model predicts $Br(T\rightarrow Ht)\leq 0.125$ in regions of parameter space consistent with electroweak precision data, this limit is inapplicable to the DESM + singlet model. CMS provides a stronger direct limit of $920$ GeV for top partners $T$ such that $Br(T\rightarrow Wb)\approx 1$ \cite{Khachatryan:2015oba}. Because the DESM + singlet model predicts a large $Br(T\rightarrow Wb)$ throughout most of the parameter space consistent with electroweak precision data, we will designate regions satisfying $m_T>920$ GeV as viable.}

{The separate panes of Figure 2 plot the viable regions of the DESM + singlet model when constrained by $\hat{T}$, $\delta g_{Lb}$, and $m_T>920$ GeV for several values of $\sin\beta$. The region of downward sloping (from left-to-right) stripes in Figure 2 is where the lightest top partner would be lighter than the physical top quark. By construction, that region is excluded, since the top partners are to be heavier than the top quark. The upward sloping (darker) striped region is where the top Yukawa coupling becomes non-perturbative; this lies outside the realm of our analysis. The coarsely gridded regions correspond to points where $\hat{T}$ lies outside of the experimentallly allowed limits $\hat{T}=0.01\pm0.12$. The more finely cross-hatched regions correspond to points where $g_{Lb}$ lies outside of the experimentally allowed values $g_{Lb}=-0.4182\pm0.0015$. The light grey curves intersecting within the white region indicate the loci of the central experimental values of $\hat{T}$ and ${g}_{Lb}$. Finally, the checkerboard black-and-gray region is excluded by the approximate direct top partner search bound, $m_T > 920$ GeV. The remaining white regions correspond to viable parts of parameter space.}

{As indicated in Figure 2, the values of $\sin\beta$ where precision electroweak tests are weakest ($\sin\beta\lesssim 0.2$) remain untouched by even the strongest limits from direct searches for top partners. As $\sin\beta$ increases, the direct searches have increasing impact.  Interestingly,  the limit ($m_T > 920$ GeV) impacts parameter space most significantly for $\sin\beta$ values where $m_T$ was already tightly constrained by precision data (as illustrated in Figure 1). For instance, the limit first makes contact with the available parameter space for $\sin\beta$ values slightly below $\sin\beta=0.3$. At $\sin\beta=0.3$, the mass of the lightest top partner ranges from $0.81\text{ TeV}$ to $2.4\text{ TeV}$. Compare this to the case at $\sin\beta=0.1$, where the limit has no impact on parameter space points that reproduce electroweak precision data, and where the lightest top partner mass ranges from $1.5\text{ TeV}$ to $17\text{ TeV}$. The mass range at $\sin\beta=0.1$ is nearly ten times as large as the mass range at $\sin\beta=0.3$. As we move toward higher values of $\sin\beta$, the range of allowed $m_T$ values contracts and moves toward lower masses. Consequently, we find that for values of $\sin\beta$ above $0.55$, there is no viable region of parameter space left.}

{The model also predicts an exotic quark $\Omega$ with an electric charge of $+5/3$ and a tree-level mass $m_\Omega = M_q$. Between ATLAS and CMS, the strongest direct limit on charge $+5/3$ quarks is $960$ GeV \cite{Aad:2015mba},\cite{CMS:2015alb}. Any point of parameter space which is excluded by this constraint (i.e., points for which $m_\Omega < 960$ GeV) is already excluded in Figure 2. So this constraint does not presently add to our understanding of the phenomenology.}

{Our analysis establishes that the DESM + singlet model remains viable for $0.05\lesssim \sin\beta \lesssim 0.55$, and strongly suggests that the full bi-doublet + singlet model is similarly viable across a significant swath of parameter space.}

\begin{figure}[h]
	\centering
	\includegraphics[scale=0.60]{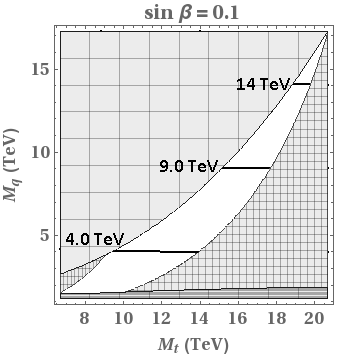}\hspace{10 pt}
	\includegraphics[scale=0.60]{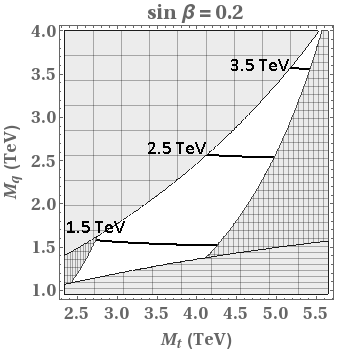}\\
	\vspace{10 pt}
	\includegraphics[scale=0.60]{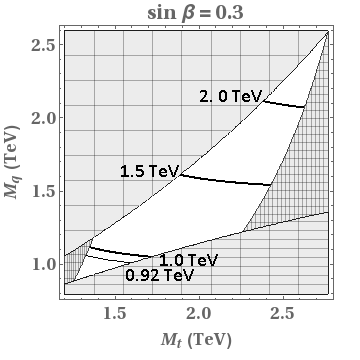}\hspace{10 pt}
	\includegraphics[scale=0.60]{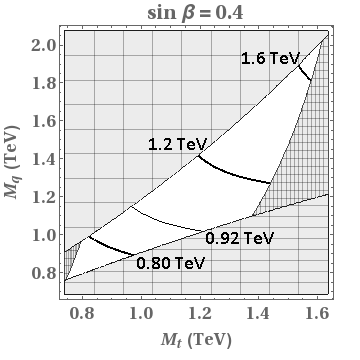}\\
	\vspace{10 pt}
	\includegraphics[scale=0.60]{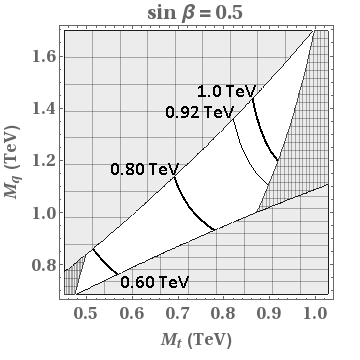}\hspace{10 pt}
	\includegraphics[scale=0.60]{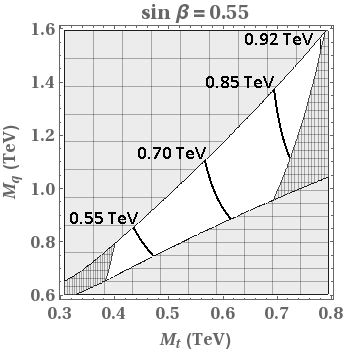}
	\captionsetup{width=0.8\textwidth}
	\caption{{Illustration of how the mass ($m_T$) of the lighter top partner varies within the region of parameter space allowed by precision EW data.  Several contours of constant $m_T$ are labeled in each plot. Each panel shows the $M_q$ vs. $M_t$ plane for a particular value of $\sin\beta$. In the white regions, the DESM + singlet model produces $\hat{T}$ and $\delta g_{Lb}$ values consistent to within $1\sigma$ of experimental bounds. We exclude the more coarsely (finely) cross-hatched regions because those points produce values of $\hat{T}$ (${g}_{Lb}$) outside the $1\sigma$ experimental band. The region excluded by $\hat{T}$ that lies above (below) the white region consists of points for which $\hat{T}$ is greater (less) than the allowed range of values. The region excluded by $g_{Lb}$ that lies left (right) of the white region consists of points for which $g_{Lb}$ is greater (less) than the allowed range of values. The thin contour in the panes for $\sin\beta \geq 0.3$ indicates the lowest value of the top partner mass now allowed by CMS direct searches; nearly all $m_T$ values in the white region for $\sin\beta = 0.55$ are below that limit. Note that the ranges of $M_t$ and $M_q$ on the axes vary between plots.}}
\end{figure}

\begin{figure}[h]
	\centering
	\includegraphics[scale=0.60]{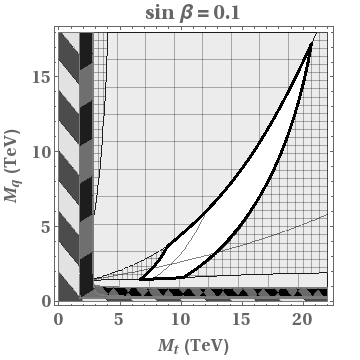}\hspace{10 pt}
	\includegraphics[scale=0.60]{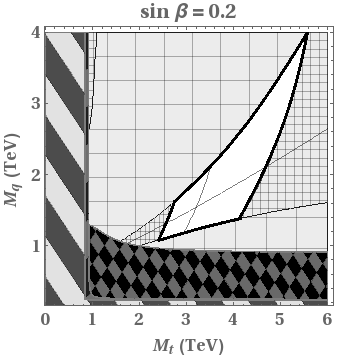}\\
	\vspace{10 pt}
	\includegraphics[scale=0.60]{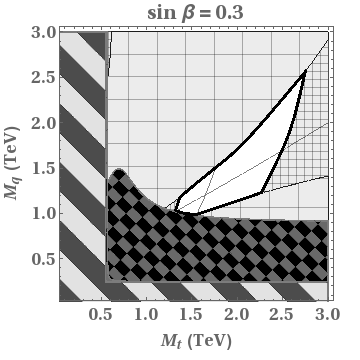}\hspace{10 pt}
	\includegraphics[scale=0.60]{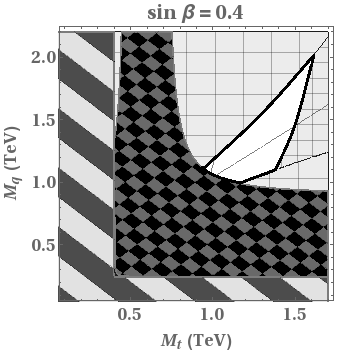}\\
	\vspace{10 pt}
	\includegraphics[scale=0.60]{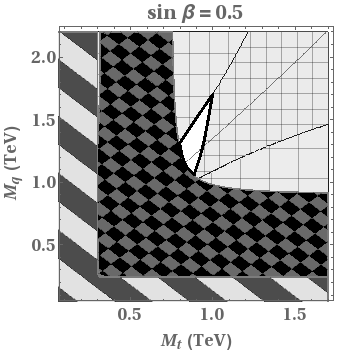}\hspace{10 pt}
	\includegraphics[scale=0.60]{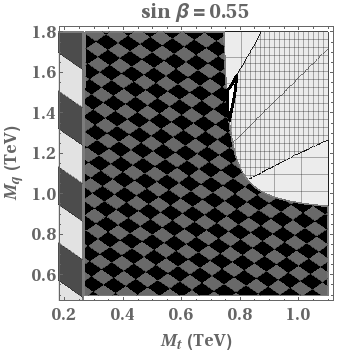}
	\captionsetup{width=0.8\textwidth}
	\caption{{The DESM + singlet model is viable within the thick-bordered white regions which survive precision EW tests and a conservative estimate of the direct search limits on top partners. Each panel shows the $M_q$ vs. $M_t$ plane for a particular value of $\sin\beta$. We exclude the L-shaped region of downward-sloping stripes by requiring the top partner masses to exceed the physical top quark mass. We exclude the L-shaped region of darker, upward-sloping stripes by keeping the Yukawa coupling of the top quark less than $4\pi$. The cross-hatched regions are excluded by precision EW data as described in Figure 1. The checkerboard black-and-gray region is excluded by CMS direct searches for top partners \cite{Khachatryan:2015oba}; points in this region produce a top partner with mass below $920$ GeV.  No viable region survives for $\sin\beta \geq 0.56$. Note that the ranges of $M_t$ and $M_q$ on the axes vary between plots.}}
\end{figure}

\section{Conclusion}

In this paper, we assessed the bi-doublet + singlet model for viability against current electroweak precision data ($\hat{T}$, $g_{Lb}$) and direct searches for top-quark partners. This model of top compositeness extends the standard model's spectrum by adding a custodially-symmetric vector-like electroweak bi-doublet and a vector-like electroweak singlet. Hence, it introduces three top partners, one bottom partner, and an exotic quark with electric charge $+5/3$. Our analysis focused on the $\mu_q\rightarrow\infty$ limit of this model, which eliminated the bottom partner and one top partner, resulting in the DESM + singlet model; we expect the phenomenology to be qualitatively similar to that of the bi-doublet + singlet model. 

{In Figure 2, we illustrate the area of model parameter space that is consistent with electroweak precision data and direct searches for top partners (the thick-bordered white region) in the $M_q$ vs. $M_t$ plane.} Overall, the DESM + singlet model (and hence the bi-doublet + singlet model) remains viable against constraints due to $\hat{T}$, $g_{Lb}$, and direct searches for top partners. 

Backovi\`{c} et. al. estimate that searches at Run-II of the LHC could be sensitive to top partner masses up to $2\text{ TeV}$ \cite{Backovic:2014}. If the bi-doublet + singlet model is a correct description of nature, a top partner discovery might well be made within the next decade. On the other hand, if no top partner is seen up to masses of 2 TeV, our conservative analysis documented in Figure 1 suggests this would eliminate parameter space points for which $\sin\beta\gtrsim 0.30$. A detailed analysis, similar to what was reported above at the end of section 4, would then be needed to establish the exact range of model parameter space that remains.

\section*{Acknowledgments}

 The work of. R.S.C., D.F., and  E.H.S. was supported by the National Science Foundation under Grants PHY-0854889 and PHY-1519045.  R.S.C. and E.H.S. also acknowledge the support of NSF Grant PHYS-1066293 and the hospitality of the Aspen Center for Physics during work on this paper.

\end{document}